\documentclass[reprint,amsmath,amssymb,aps]{revtex4-2}
\newcommand{\be}{\begin{equation} } 
\newcommand{\ee}{\end{equation} } 
\newcommand{\ba}{\begin{array} } 
\newcommand{\ea}{\end{array} } 
\newcommand{\bear}{\begin{eqnarray} } 
\newcommand{\eear}{\end{eqnarray} }

\newcommand{\sev}{S_{\rm ev}}
\newcommand{\bev}{B_{\rm ev}}

\usepackage{graphicx, comment, multirow, appendix}
\usepackage{dcolumn}
\usepackage{bm}
\usepackage{hyperref}
\usepackage{lineno}
\usepackage{float}
\usepackage{cleveref}

\begin{document}

\noindent \makebox[5.cm][l]{\small \hspace*{-.2cm} }{\small IFIN-DFPE-26-0011}  
\title{HL-LHC sensitivity to an ultraheavy $S_{uu}$ diquark in the $u\chi$ channel}
\author{Matei S. Filip$^{1,2}$, Calin Alexa$^1$, Daniel C. Costache$^{1,2}$, Ioan M. Dinu$^1$, Ioana Duminica$^{1,2}$ and Gabriel C. Majeri$^2$}
\affiliation{
$^1$Particle Physics Department, IFIN-HH, 077125 M\u agurele, RO \\
$^2$Faculty of Physics, University of Bucharest, 077125 M\u agurele, RO }
\date{\today}
\email{Corresponding author: calin.alexa@cern.ch}
\begin{abstract}
We study the HL-LHC sensitivity to an ultraheavy diquark $S_{uu}$ produced in up-quark fusion and decaying as $S_{uu} \rightarrow u\chi$, $\chi \rightarrow Wb, \ Zt, \ h^0t$. For fully hadronic decays of the W, Z and top quark, this gives rise to multijet final states. Within the same model framework used previously for the $S_{uu}\to\chi\chi$ six-jet channel, we consider $S_{uu}$ masses in the multi-TeV range and vectorlike quark masses of order a few TeV, and simulate proton–proton collisions at $\sqrt s = 13.6$ TeV with integrated luminosities up to the HL-LHC target. The analysis strategy employs a machine-learning–based discriminant adapted from the six-jet study to the new four-jet topology, which we use to derive the corresponding discovery reaches and exclusion limits. We find that this topology improves the overall sensitivity to $S_{uu}$ in regions where the branching ratio $B(S_{uu}\to u\chi)$ is sizable and provides a complementary signature for studying ultraheavy diquarks at the HL-LHC.
\end{abstract}
\maketitle

\section{\label{sec:intro}Introduction}

The search for heavy resonances at the TeV scale remains an important objective of the ATLAS and CMS experiments at the High Luminosity Large Hadron Collider (HL-LHC). Multijet final states have been studied in searches for a wide range of Beyond the Standard Model (BSM) scenarios, including color-sextet diquark scalars and vectorlike quarks (VLQs) ~\cite{CMS:2025hpa,CMS:2022usq,ATLAS:2023ssk,ATLAS:2024gyc}. Recent CMS searches for ultraheavy resonances in four-jet final states~\cite{CMS:2025hpa} have shown sensitivity to masses up to 9 TeV, suggesting that the forthcoming HL-LHC will significantly extend the discovery potential for new colored sextet diquarks. Hints of excesses in di-(di)jet searches at the LHC are discussed in \cite{Crivellin:2022nms}.

One such BSM framework \cite{Dobrescu:2018psr,Dobrescu:2019nys,Dobrescu:2024mdl} features a color-sextet scalar diquark $S_{uu}$ with Yukawa couplings $y_{uu}$ to right up-type quarks, $y_{u\chi}$ to $u\chi$, and $y_{\chi\chi}$ to same-sign vectorlike quarks $\chi$. The $S_{uu}$ particle is a complex scalar with electric charge $+4/3$, while the VLQ is a color-triplet fermion of charge $+2/3$. Depending on the mass hierarchy, the scalar may decay either into a pair of VLQs ($S_{uu}\to \chi\chi$) or into a single VLQ and an up quark ($S_{uu}\to u\chi$). Both channels yield distinctive signatures at hadron colliders due to their high jet multiplicity and same-sign content.

In our previous works \cite{Duminica_2025}, we analyzed the six-jet final state $S_{uu} \to \chi\chi \to (Wb)(Wb) \to (jjb)(jjb)$, providing a detailed Machine Learning (ML) based signal selection and a comprehensive statistical evaluation of the discovery and exclusion prospects at the HL-LHC \cite{Costache_2025}.

The present study complements those analyses by exploring the single-VLQ production channel $S_{uu}\to u\chi$, where the VLQ decays as $\chi\to Wb,~Zt,~h^0t$, considering all possible combinations of fully hadronic final states. This process results in a set of multijet final states with high sensitivity potential at the HL-LHC.

We assess the sensitivity of this channel using Machine Learning–based signal selection and a comprehensive statistical framework. This complementary analysis improves the sensitivity to $S_{uu}$ production and establishes a consistent framework for future combined analyses.

The paper is organized as follows: \Cref{sec:TH-Data} briefly reviews the BSM framework, the Monte Carlo (MC) generation of data samples and the Machine Learning method for signal–background discrimination. Selected signal and background events are presented in \Cref{sec:results}, including discussions regarding the impact of process parameters on our Random Forest (RF) based analysis. The statistical framework and results are comprehensively treated in \Cref{sec:statistics}. Finally, \Cref{sec:conclusions} summarizes the main findings and future prospects.

\section{\label{sec:TH-Data}Theoretical framework and signal selection}

The theoretical model underlying this study is the same color-sextet scalar diquark framework introduced in Refs.~\cite{Dobrescu:2018psr,Dobrescu:2019nys,Dobrescu:2024mdl}. The present work focuses exclusively on the $S_{uu}\to u\chi$ decay mode with $\chi$ undergoing fully hadronic decays through the $\chi\to Wb, \chi\to Zt$ and $\chi\to h^0t$ channels.

The tree-level partial decay width of the diquark scalar into an up quark and a vectorlike quark is \cite{Dobrescu:2018psr}:
\begin{equation}
    \Gamma\left(S_{uu} \rightarrow u \chi\right) = \frac{M_S}{16 \pi} \, |y_{u \chi}|^2 \left(1-\frac{m_\chi^2}{M_S^2} \right)^2
    \label{eq:BR}
\end{equation}
where $M_S$ is the $S_{uu}$ mass and $m_\chi$ is the VLQ mass. This expression assumes a two-body decay at tree level in the narrow-width approximation, neglects the up-quark mass, and uses a scalar–fermion–fermion coupling. For $m\chi=2~\text{TeV}$, the branching ratio $B(S_{uu}\to u\chi)$ is $14.2\%$ for $M_S=7~\text{TeV}$ and $13.7\%$ for $M_S=10~\text{TeV}$.

The Monte Carlo (MC) samples were generated using \textsc{MadGraph5\_aMC@NLO} \cite{Alwall:2014hca} interfaced with \textsc{Pythia8.310} \cite{Bierlich:2022pfr} for parton showering and hadronization, using the NNPDF23LO~\cite{NNPDF:2014otw} parton distribution functions. Detector response was simulated using \textsc{Delphes 3.5.0}~\cite{deFavereau:2013fsa} with the ATLAS detector default card. Jet reconstruction employs the anti-$k_t$ algorithm as implemented in \textsc{FastJet}~\cite{Cacciari:2011ma}. In both signal and data samples, initial (ISR) and final state radiation (FSR), alongside multi-parton interactions (MPI) are handled by \textsc{Pythia8.310} \cite{Bierlich:2022pfr}. We considered the same background processes as in Ref.~\cite{Duminica_2025} and ran them through an identical simulation chain for consistency and validation. We generate signal samples for $7 ~\text{TeV}\le M_S\le 10~\text{TeV}$ with $m_\chi = 1.5~\text{TeV}$ and $2~\text{TeV}$. 

\begin{figure}[h]
    \centering
    \includegraphics[width=0.5\textwidth]{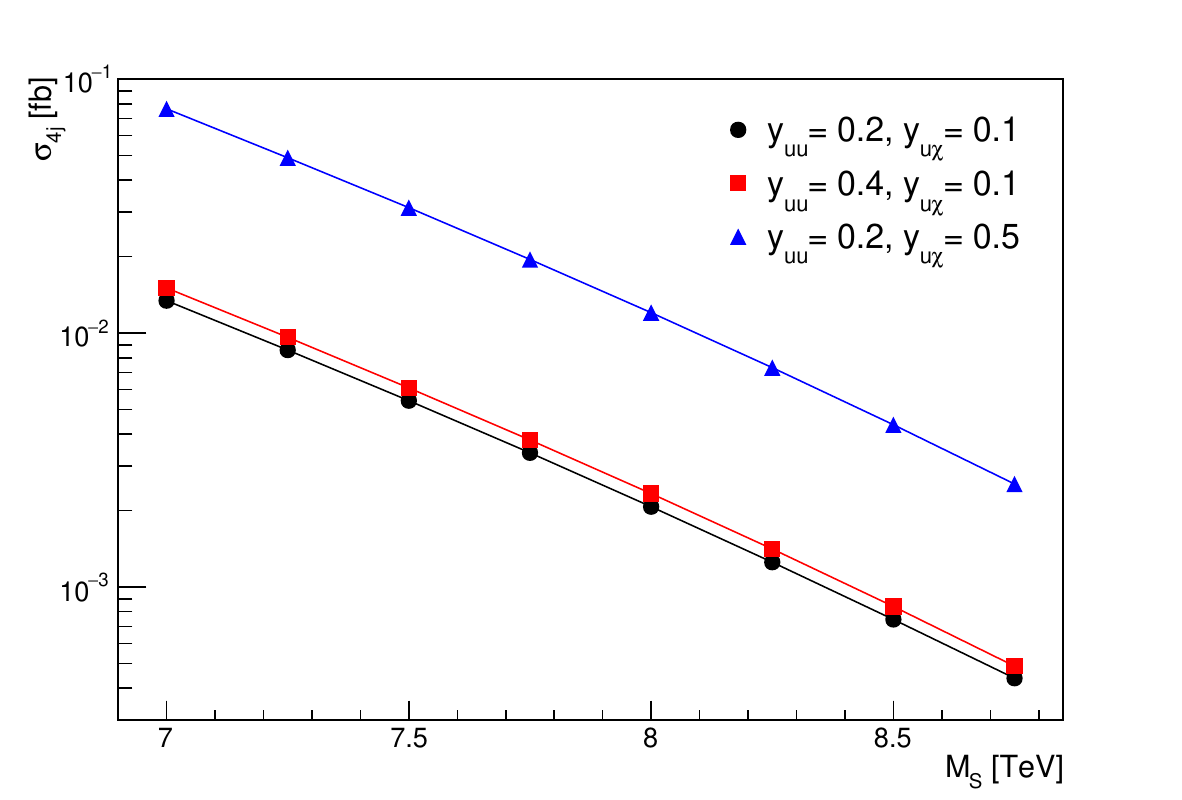}
    \caption{$pp\rightarrow S_{uu}\rightarrow u\chi\rightarrow u(Wb)\rightarrow u(jjb)$ cross section as a function of $M_S$ and $y_{uu}$, $y_{u\chi}$ at $\sqrt{s}=13.6$ TeV.}
    \label{fig:Crossx_uChi}
\end{figure}

We consider three benchmark coupling scenarios. In the first two, following our previous studies, the product $y_{uu}B(\chi\to Wb)$ is set to 0.1 and 0.2, with $y_{u\chi}=0.1$. The third scenario explores the possibility that the $u\chi$ channel dominates $S_{uu}$ discovery, with an enhanced Yukawa coupling $y_{u\chi}=0.5$. \Cref{fig:Crossx_uChi} shows the dependence of the four-jet cross section on the Yukawa couplings $y_{uu}$ and $y_{u\chi}$. For fixed $y_{u\chi}$, increasing $y_{uu}$ changes the cross section by about 11\%. In comparison, raising $y_{u\chi}$ to 0.5 leads to a much larger effect, with the four-jet cross section enhanced by a factor of about 6.

Studies done by both ATLAS~\cite{ATLAS:2024gyc} and CMS~\cite{CMS:2022fck} Collaborations suggest that same-sign vectorlike quarks mass lower bounds are set at 1.36 TeV and 1.4 TeV, respectively, in pair production processes. \Cref{tab:crossx_VLQ_masses} shows that when lowering the vectorlike quark mass from 2 TeV to 1.5 TeV, the cross section drops with only $\approx 2.5 \%$.

\begin{ruledtabular}
\begin{table}[h]
\caption{\label{tab:crossx_VLQ_masses}
$pp \rightarrow S_{uu} \rightarrow u\chi \rightarrow u(Wb) \rightarrow u(jjb)$ process cross section vs $M_S$ for $m_\chi=1.5,~2$ TeV at $\sqrt{s}=13.6$ TeV, in the conservative $y_{uu}=0.2$, $y_{u\chi}=0.1$ scenario.}
\centering
\footnotesize
\begin{tabular}{l|cccc}
& $M_S=7$ TeV & $M_S=7.5$ TeV & $M_S=8$ TeV & $M_S=8.5$ TeV  \\[1mm]   
\hline
\\[-1.6mm]
\multicolumn{5}{c}{$m_\chi=2$ TeV} \\[1mm]
\hline \\[-2.6mm]
$\sigma_{4j}$[fb] & 1.34E-02 & 5.42E-03 & 2.07E-03 & 7.46E-04\\[1mm]
\hline
\\[-1.6mm]
\multicolumn{5}{c}{$m_\chi=1.5$ TeV} \\[1mm]
\hline \\[-2.6mm]
$\sigma_{4j}$[fb] & 1.31E-02 & 5.30E-03 & 2.04E-03 & 7.34E-04\\[1mm]
\end{tabular}
\end{table}
\end{ruledtabular}

\begin{figure}[h]
    \centering
    \includegraphics[width=0.5\textwidth]{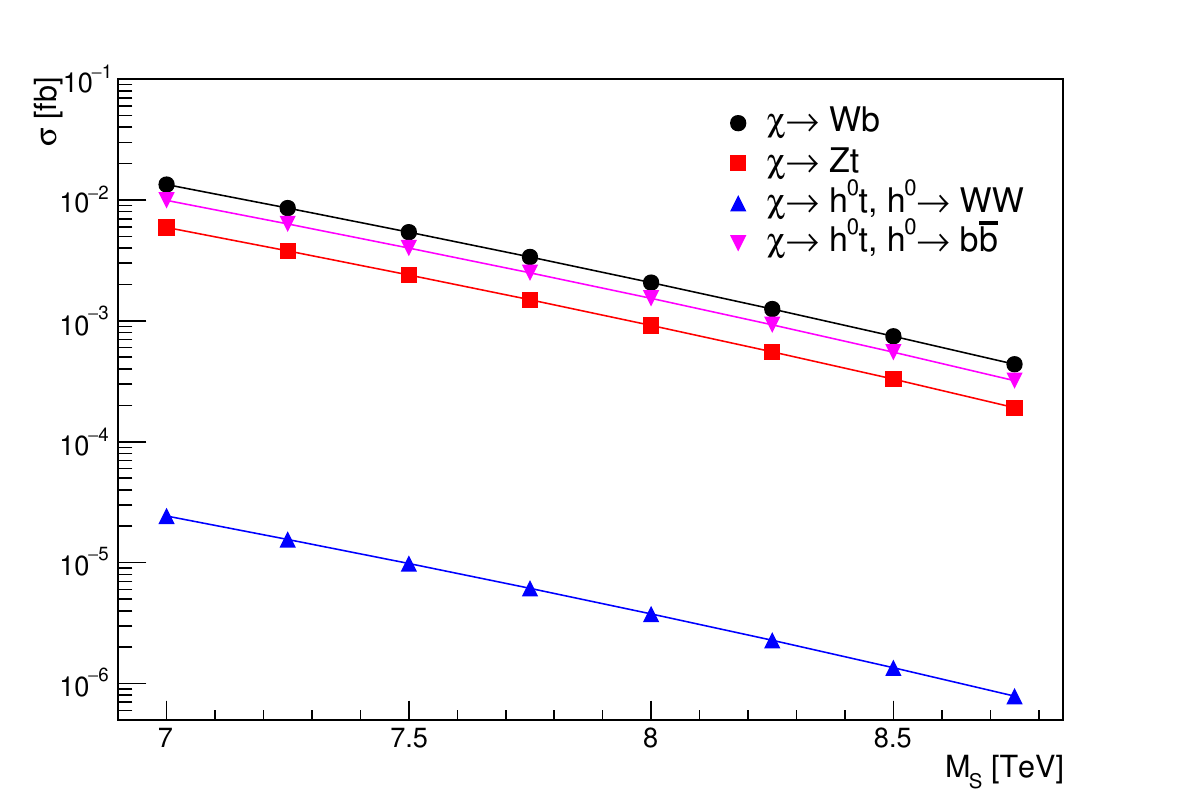}
    \caption{
    Fully hadronic cross sections for different $\chi$ decay modes in $S_{uu}\rightarrow u\chi$ at $\sqrt{s}=13.6$ TeV, for $y_{uu}=0.2$ and $y_{u\chi}=0.1$.}
    \label{fig:Crossx_uChi_decay_channels}
\end{figure}

While the CMS studies \cite{CMS:2025hpa,CMS:2022usq} target the scenario of $B(\chi\rightarrow jj)=100\%$, we instead consider three different vectorlike quark decay modes, with the corresponding branching ratios: $B(\chi \rightarrow W b$):$B(\chi \rightarrow h^0 t$):$B(\chi \rightarrow Z t ) =$ $50\%$:$25\%$:$25\%$. The cross sections are shown in \Cref{fig:Crossx_uChi_decay_channels} assuming $t\to Wb$ and fully hadronic decays of the W and Z bosons. In the following sections, we neglect the $\chi \rightarrow h^0t$, $h^0\rightarrow WW$ channel, whose cross section is several orders of magnitude smaller than the others.

The present analysis introduces a new production topology, $S_{uu}\to u\chi$, which yields a distinct final-state configuration and different kinematic distributions. This provides an additional, complementary way to constrain the $S_{uu}$ model and improves the global sensitivity when combined with the previous results.

MC-generated signal and background samples are subsequently processed with a Machine Learning classifier optimized for signal-background discrimination. The same Random Forest algorithm framework developed in Ref.~\cite{Duminica_2025} is employed here, with additional optimization for the kinematic features characteristic of the $S_{uu}\to u\chi$ topology. The output of the classifier serves as the primary discriminant in the statistical analysis, which follows the same methodological principles as in our previous study but is adapted to the present final state and event composition.

\section{\label{sec:results} Signal and background classification}

Signal–background separation is performed with a Random Forest classifier trained on various $S_{uu}\to u\chi$ scenarios. We examine how the vectorlike quark mass and the Yukawa couplings influence the signal selection event counts for all $\chi$ decay modes.

\subsection{\label{subsec:VLQ_mass}Vectorlike quark mass}

As discussed before, lowering the vectorlike quark mass from 2 to 1.5~TeV has only a small effect on the four-jet cross section, with $B(S_{uu}\to u\chi)$ decreasing slightly from 13.9\% to 13.7\%. 
The corresponding signal $(S_{ev})$ and background $(B_{ev})$ event yields for both mass points, for discriminator thresholds $D\in \left[0.9, 0.97\right]$, are shown in \Cref{tab:VLQ_masses}. In line with the small change in cross section, $S_{ev}$ is slightly lower for the lighter $\chi$, while the background event yields vary only within statistical uncertainties.

\begin{ruledtabular}
\begin{table}[h]
\caption{\label{tab:VLQ_masses}
Random Forest selected signal ($\sev$) and background ($\bev$) event yields for $S_{uu} \rightarrow u \chi \rightarrow u (Wb)$ with $m_\chi =1.5, ~2$ TeV, $M_S=8$ TeV and $\widehat{m}_{\text{min}}=7.5$ TeV.
}
\centering
\footnotesize
\begin{tabular}{l|ccccc}
& $D=0.90$ & $D=0.925$ & $D=0.95$ & $D=0.96$ & $D=0.97$ \\[1mm]   
\cline{2-6}
\\[-1.6mm]
\multicolumn{6}{c}{$m_\chi=2$ TeV} \\[1mm]
\hline \\[-2.6mm]
$\sev$ & 6.04 & 5.91 & 5.40 & 4.85 & 4.00 \\[1mm]
$\bev$ & 3.15{\tiny$\pm$}0.26 & 1.88{\tiny$\pm$}0.42 & 0.72{\tiny$\pm$}0.19 & 0.44{\tiny$\pm$}0.23 & 0.17{\tiny$\pm$}0.06 \\[1mm] \hline \\[-2.6mm]
\multicolumn{6}{c}{$m_\chi=1.5$ TeV} \\[1mm]
\hline \\[-2.6mm]
$\sev$ & 5.89 & 5.70 & 5.10 & 4.59 & 3.82 \\[1mm]
$\bev$ & 3.26{\tiny$\pm$}0.60 & 1.99{\tiny$\pm$}0.48 & 0.78{\tiny$\pm$}0.21 & 0.46{\tiny$\pm$}0.18 & 0.25{\tiny$\pm$}0.21 \\[1mm]
\end{tabular}
\end{table}
\end{ruledtabular}

\subsection{\label{subsec:yukawa}Conservative and discovery scenarios with Yukawa couplings}

We consider three coupling scenarios for the fully hadronic $u\chi$ final state: (i) a conservative choice $y_{uu}=0.2$, $ y_{u\chi}=0.1$; (ii) an enhanced diquark coupling $y_{uu}=0.4$ with $y_{u\chi}=0.1$; (iii) a scenario in which $u\chi$ channel is favoured, with $y_{u\chi}=0.5$, as motivated in Refs.\cite{Dobrescu:2018psr,Dobrescu:2019nys,Dobrescu:2024mdl}. The corresponding signal event yields $S_{ev}$ are summarized in \Cref{tab:yukawa_couplings}, with the largest $S_{ev}$ obtained for $y_{u\chi}=0.5$ and the smallest $S_{ev}$ in the conservative scenario.

\begin{ruledtabular}
\begin{table}[h]
\caption{\label{tab:yukawa_couplings}
Sampled signal and background event counts obtained with the Random Forest model for the fully hadronic $S_{uu} \rightarrow u \chi \rightarrow u (Wb)$ channel, $m_\chi = 2$ TeV,  $M_S=8$ TeV and $\widehat{m}_{\text{min}}=7.5$ TeV in 3 different Yukawa scenarios.
}
\centering
\footnotesize
\begin{tabular}{l|ccccc}
& $D=0.90$ & $D=0.925$ & $D=0.95$ & $D=0.96$ & $D=0.97$ \\[1mm]   
\cline{2-6}
\\[-1.6mm]
\multicolumn{6}{c}{$y_{uu}B(\chi\rightarrow Wb) = 0.1$, $y_{u \chi}=0.1$} \\[1mm]
\hline \\[-2.6mm]
$\sev$ & 6.04 & 5.91 & 5.40 & 4.85 & 4.00 \\[1mm]
$\bev$ & 3.15{\tiny$\pm$}0.26 & 1.88{\tiny$\pm$}0.42 & 0.72{\tiny$\pm$}0.19 & 0.44{\tiny$\pm$}0.23 & 0.17{\tiny$\pm$}0.06 \\[1mm]
\hline \\[-1.6mm]
\multicolumn{6}{c}{$y_{uu}B(\chi\rightarrow Wb) = 0.2$, $y_{u \chi}=0.1$} \\[1mm]
\hline \\[-2.6mm]
$\sev$ & 6.80 & 6.65 & 6.06 & 5.49 & 4.52 \\[1mm]
$\bev$ & 3.11{\tiny$\pm$}0.43 & 1.83{\tiny$\pm$}0.42 & 0.76{\tiny$\pm$}0.19 & 0.44{\tiny$\pm$}0.21 & 0.16{\tiny$\pm$}0.07 
\\[1mm]
\hline \\[-1.6mm]
\multicolumn{6}{c}{$y_{uu}B(\chi\rightarrow Wb) = 0.1$, $y_{u \chi}=0.5$} \\[1mm]
\hline \\[-2.6mm]
$\sev$ & 35.10 & 34.28 & 31.25 & 28.28 & 23.40 \\[1mm]
$\bev$ & 3.22{\tiny$\pm$}0.34 & 1.83{\tiny$\pm$}0.41 & 0.72{\tiny$\pm$}0.33 & 0.34{\tiny$\pm$}0.15 & 0.17{\tiny$\pm$}0.06
\end{tabular}
\end{table}
\end{ruledtabular}

\subsection{\label{subsec:decay_channels}Decay channels}
For a comprehensive study of fully hadronic final states in the $S_{uu}\to u\chi$ channel, in addition to the $\chi\to Wb$ mode, we also include the decays $\chi\to Zt$ and $\chi\to h^0 t$. 

As shown in \Cref{tab:chi_decays}, the $\chi\to Wb$ fully hadronic mode yields the highest signal-to-background ratio, driven by its larger branching ratio fraction and production cross section relative to the other channels. 

Nevertheless, despite the different multiplicities, the Random Forest achieves similar background-rejection efficiencies for all three channels, within the cross-validation uncertainties.

\begin{ruledtabular}
\begin{table}[hb]
\caption{\label{tab:chi_decays}
Sampled signal ($\sev$) and background ($\bev$) event counts obtained with the Random Forest model for 3 $\chi$ decay channels, $y_{uu}B(\chi\rightarrow Wb) = 0.1$, $m_\chi = 2$ TeV,  $M_S=8$ TeV and $\widehat{m}_{\text{min}}=7.5$ TeV.}
\centering
\footnotesize
\begin{tabular}{l|ccccc}
& $D=0.90$ & $D=0.925$ & $D=0.95$ & $D=0.96$ & $D=0.97$ \\[1mm]   
\cline{2-6}
\\[-1.6mm]
\multicolumn{6}{c}{$S_{uu} \rightarrow u \chi \rightarrow u (Wb)$} \\[1mm]
\hline \\[-2.6mm]
$\sev$ & 6.04 & 5.91 & 5.40 & 4.86 & 4.00 \\[1mm]
$\bev$ & 3.15{\tiny$\pm$}0.26 & 1.89{\tiny$\pm$}0.42 & 0.72{\tiny$\pm$}0.19 & 0.44{\tiny$\pm$}0.23 & 0.17{\tiny$\pm$}0.06 \\[1mm]
\hline \\[-1.6mm]
\multicolumn{6}{c}{$S_{uu} \rightarrow u \chi \rightarrow u (h^0t), \ h^0 \rightarrow b \overline{b}$} \\[1mm]
\hline \\[-2.6mm]
$\sev$ & 4.53 & 4.45 & 4.12 & 3.75 & 3.13 \\[1mm]
$\bev$ & 3.43{\tiny$\pm$}0.54 & 1.84{\tiny$\pm$}0.44 & 0.56{\tiny$\pm$}0.19 & 0.31{\tiny$\pm$}0.10 & 0.18{\tiny$\pm$}0.09 \\[1mm]
\hline \\[-1.6mm]
\multicolumn{6}{c}{$S_{uu} \rightarrow u \chi \rightarrow u (Zt)$} \\[1mm]
\hline \\[-2.6mm]
$\sev$ & 2.67 & 2.60 & 2.37 & 2.14 & 1.79 \\[1mm]
$\bev$ & 3.23{\tiny$\pm$}0.58 & 1.72{\tiny$\pm$}0.42 & 0.68{\tiny$\pm$}0.23 & 0.35{\tiny$\pm$}0.10 & 0.21{\tiny$\pm$}0.09 \\[1mm]
\end{tabular}
\end{table}
\end{ruledtabular}

Taken together, these studies indicate that the search sensitivity depends more strongly on the Yukawa couplings and the $\chi$ decay mode than on the precise value of $m_\chi$ in the range studied.

This motivates using the fully hadronic $\chi\to Wb$ mode and the higher $y_{u\chi}$ scenario as representative benchmarks in the subsequent analysis.

\section{\label{sec:statistics}Statistical interpretation}

To quantify the observation potential of the resonant $S_{uu}$ production in the fully hadronic $u\chi$ channel, we follow the statistical procedure in Ref. \cite{Costache_2025}. Using the \textsc{RooFit} \cite{roofit} and  \textsc{RooStats} \cite{roostatsproject} frameworks, we compute the local $p$-values, $CL_{s}$ metrics, and the $95\%$ C.L. upper limits for each scenario discussed in \Cref{sec:results}.

\begin{figure}[h]
\centering
\includegraphics[width=0.5\textwidth]{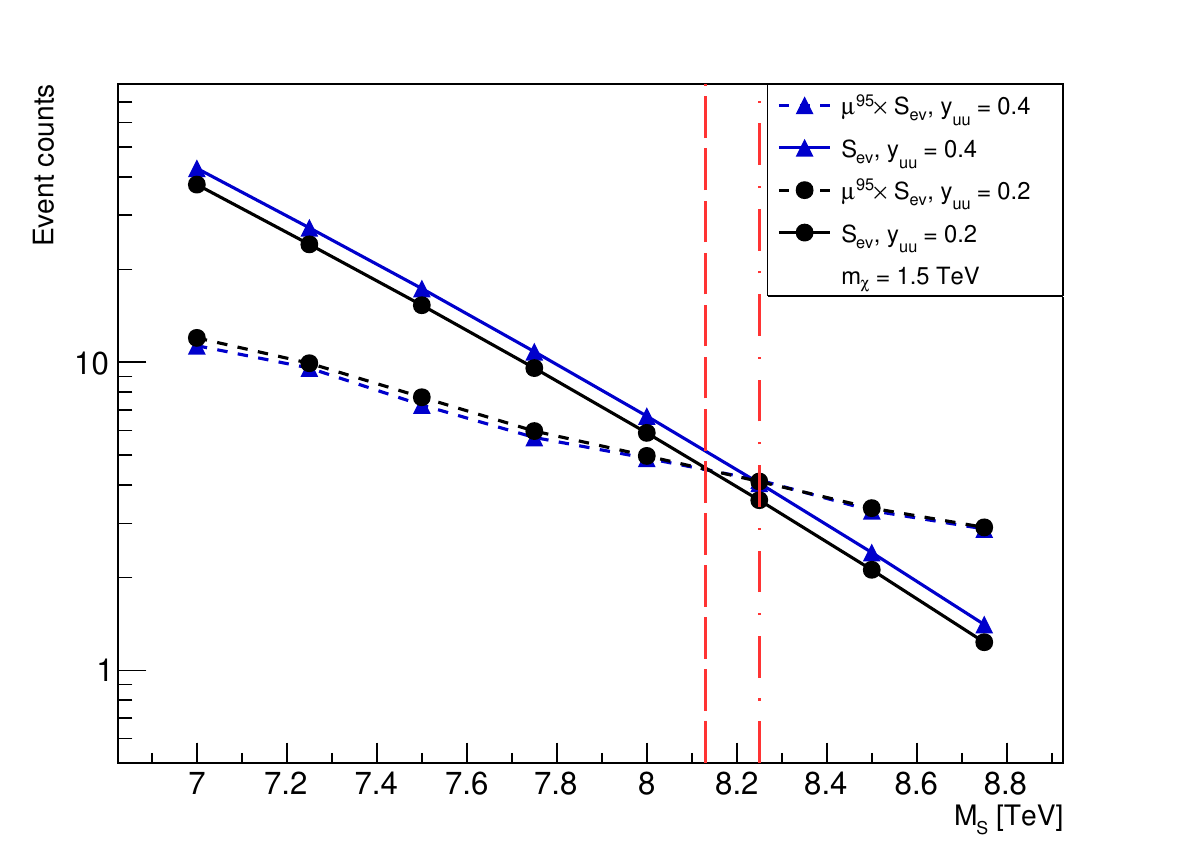} 
\caption{95\% C.L.\ Upper Limit on the signal strength multiplier $\mu$ times the ML event yield $S_{ev}$ for $S_{uu} \rightarrow u \chi \rightarrow u (Wb)$ fully hadronic channel with $m_\chi = 1.5$ TeV and $D=0.9$.}
\label{fig:chi15-ul}
\end{figure}

\begin{figure}[h]
\centering
\includegraphics[width=0.5\textwidth]{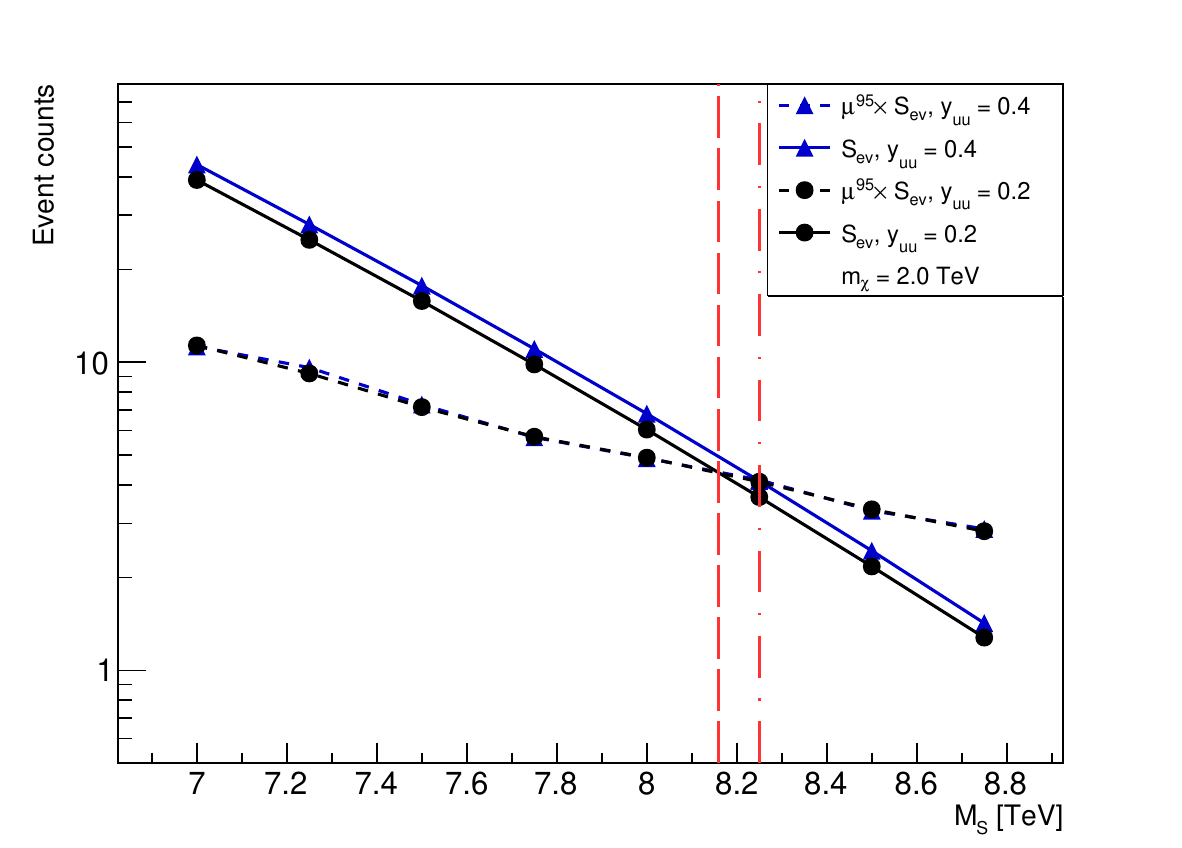} 
\caption{95\% C.L.\ Upper Limit on the signal strength multiplier $\mu$ times the ML event yield $S_{ev}$ for $S_{uu} \rightarrow u \chi \rightarrow u (Wb)$ fully hadronic channel with $m_\chi = 2$ TeV and $D=0.9$.}
\label{fig:chi2-ul}
\end{figure}

In contrast to the $\chi\chi$ channel, the expected limits in the $u\chi$ channel only change mildly when the Yukawa couplings and the VLQ mass are varied (\Cref{fig:chi15-ul,fig:chi2-ul}).

The analysis is performed for a fixed ML discriminator threshold $D=0.9$. Compared to the previous decay channel, the exclusion limits in the $u\chi$ case are clustered around $M_S\simeq 8.2~\text{TeV}$, with variations of at most $0.1~\text{TeV}$, as shown in \Cref{fig:chi15-ul,fig:chi2-ul}. These values indicate that varying $m_\chi$ and $y_{uu}$ does not significantly improve the search sensitivity, and the model predictions for the $u\chi$ decay channel are relatively stable under these parameter changes.

\begin{figure}[h]
\centering
\includegraphics[width=0.5\textwidth]{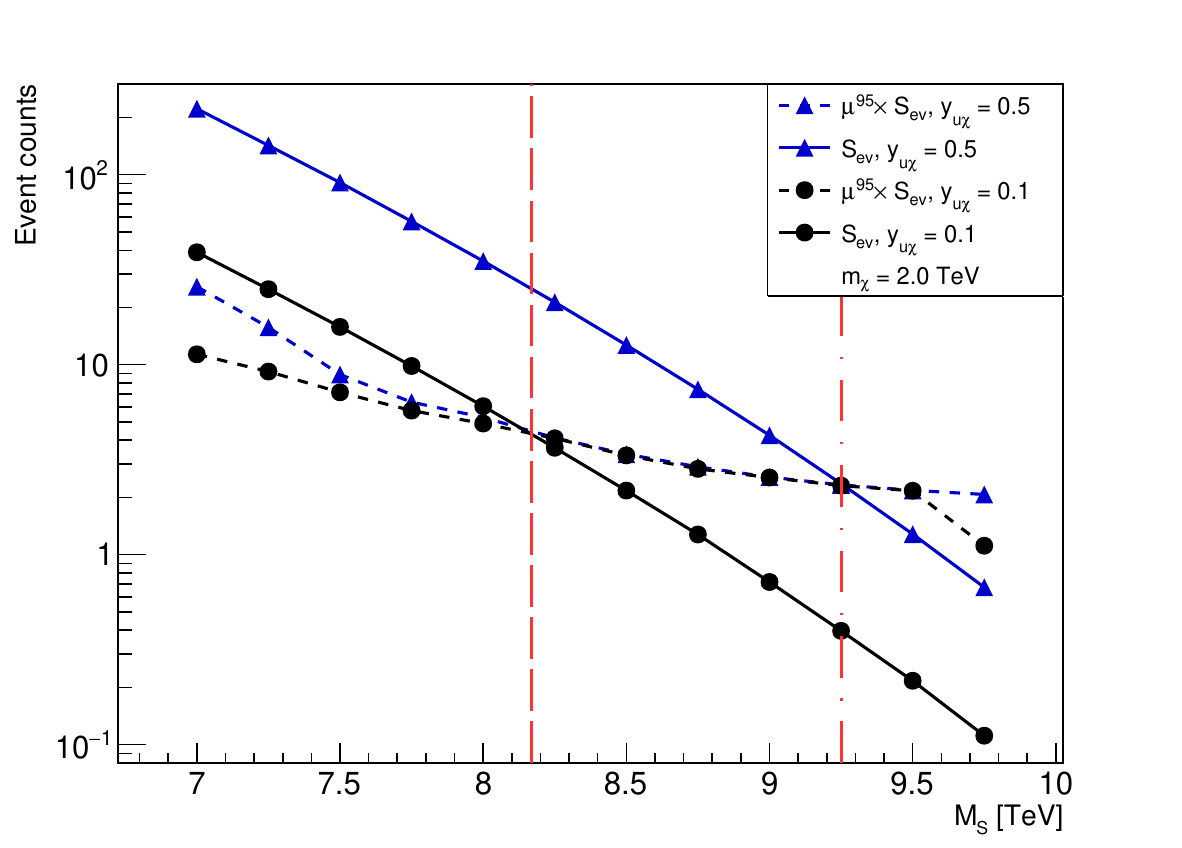} 
\caption{95\% C.L. Upper Limit on the signal strength multiplier $\mu$ times the ML event yield $S_{ev}$ for $S_{uu} \rightarrow u \chi \rightarrow u (Wb)$ fully hadronic channel for the conservative $y_{u\chi}=0.1$ case and the $y_{u\chi}=0.5$ discovery channel.}
\label{fig:yuChi-ul}
\end{figure}

By contrast, in the benchmark with enhanced coupling $y_{u\chi}=0.5$, where the $u\chi$ mode acts as the main discovery channel, the mass reach improves significantly. As shown in \Cref{fig:yuChi-ul}, the 95\% C.L. limits extend by about 1 TeV, up to $M_S=9.25~\text{TeV}$.

\begin{figure}[h]
\centering
\includegraphics[width=0.5\textwidth]{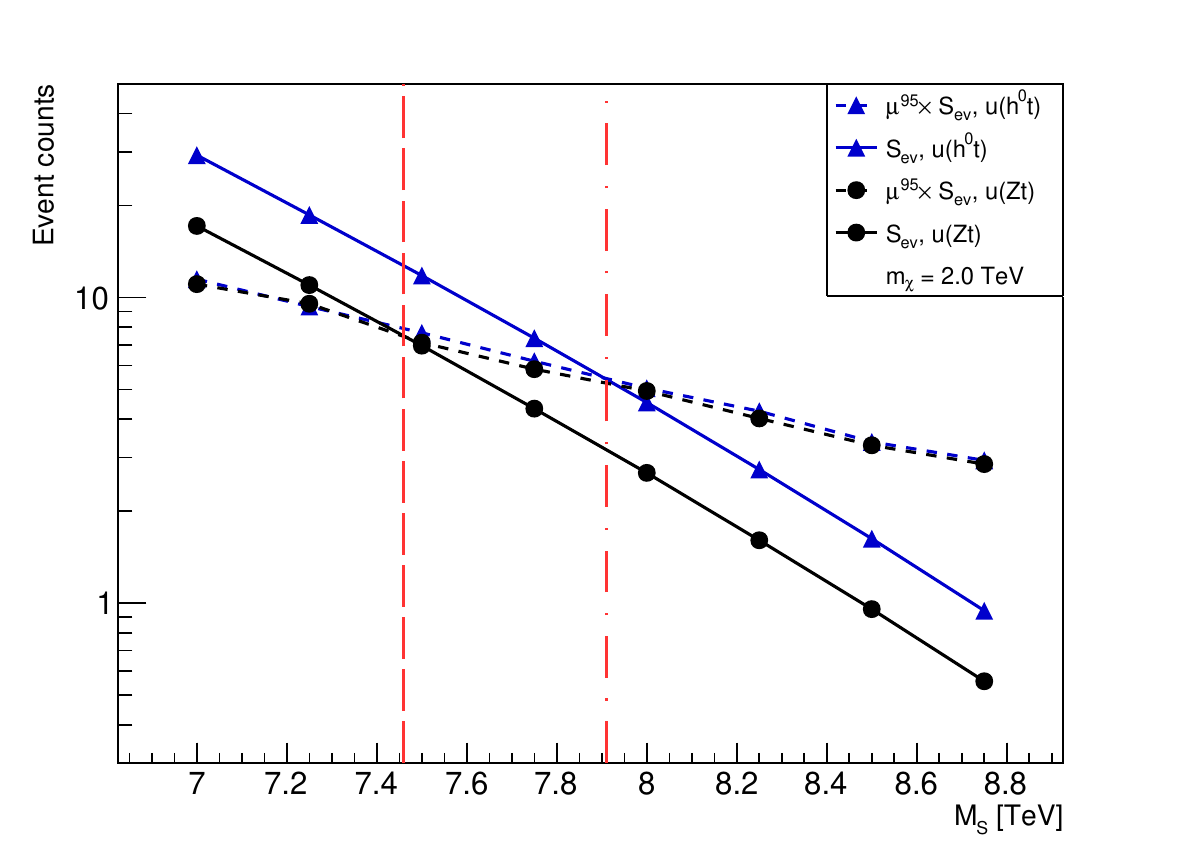} 
\caption{95\% C.L. Upper Limit on the signal strength multiplier $\mu$ times the ML event yield $S_{ev}$ for $S_{uu} \rightarrow u \chi \rightarrow u (h^0t)$ and $S_{uu} \rightarrow u \chi \rightarrow u (Zt)$ at $D=0.9$.}
\label{fig:new-chan-ul}
\end{figure}

The additional decay channels $\chi\to Zt$ and $\chi\to h^0t$ are shown in the \Cref{fig:new-chan-ul}. The scans are performed for $D=0.9$ and $y_{uu}=0.2$. In these cases, the mass reaches are reduced, with the observed intersections at $M_S\approx 7.45~\text{TeV}$ for $\chi\to Zt$ and $M_S\approx 7.9~\text{TeV}$ for $\chi\to h^0t$.

\begin{figure}[h]
\centering
\includegraphics[width=0.5\textwidth]{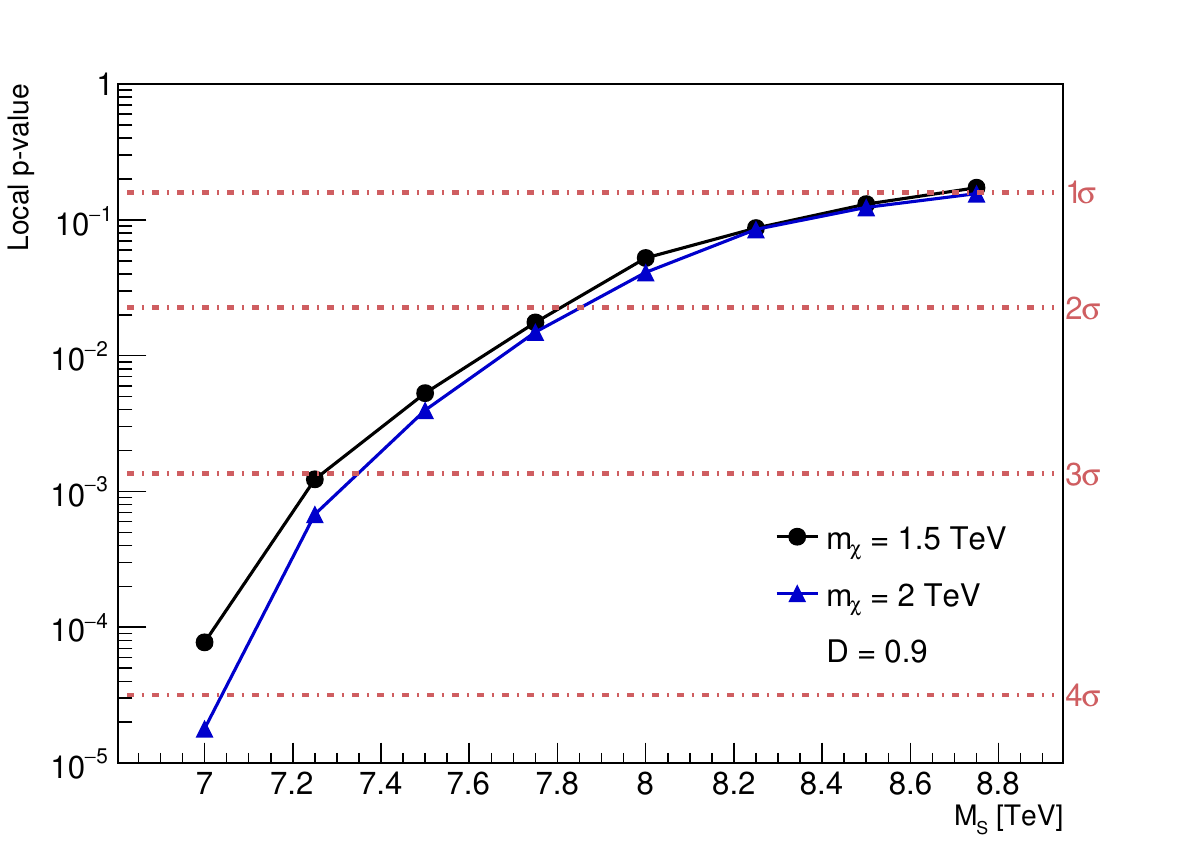} 
\caption{Local $p$-value quantifying an excess above the observed test statistic $q_{0}^{\text{obs}}$, for the two analyzed values of $m_{\chi}$, in the $S_{uu} \rightarrow u \chi \rightarrow u (Wb)$ fully hadronic channel, at $y_{uu}=0.2$.}
\label{fig:p-val}
\end{figure}

We also compute the local $p$-values, fixing the ML discriminator to $D=0.9$. In \Cref{fig:p-val} we compare the results for two VLQ mass hypotheses, $m_{\chi}=1.5~\text{TeV}$ and $m_{\chi}=2~\text{TeV}$. Unlike in our previous study \cite{Costache_2025}, the $p$-values are slightly smaller for $m_{\chi}=2~\text{TeV}$, indicating a modestly better discovery potential at the higher VLQ mass. At lower $S_{uu}$ masses, $p_0$ reaches the $\sim 4\sigma$ level, with both $m_{\chi}$ choices giving very similar results.

\begin{figure}[h]
\centering
\includegraphics[width=0.5\textwidth]{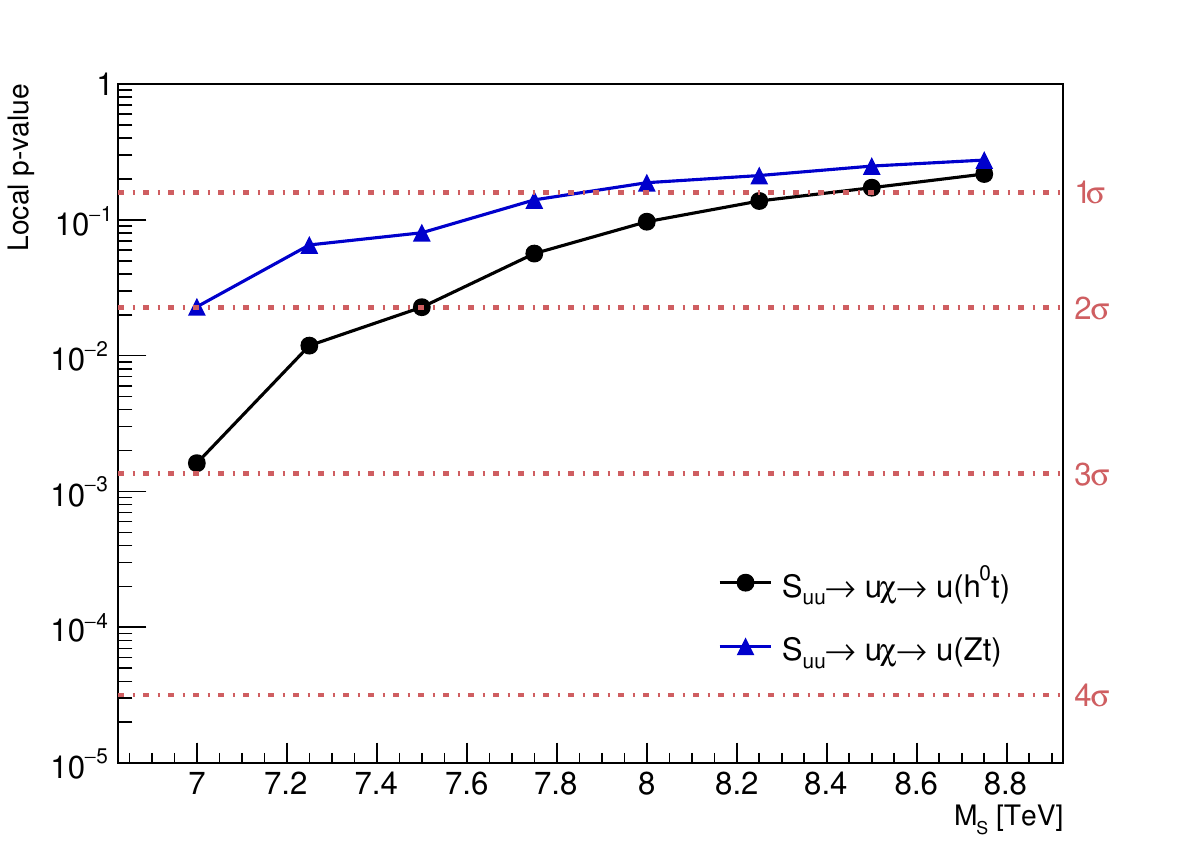} 
\caption{Local $p$-value quantifying an excess above the observed test statistic $q_{0}^{\text{obs}}$, in the $S_{uu} \rightarrow u \chi \rightarrow u (h^0t)$ and $S_{uu} \rightarrow u \chi \rightarrow u (Zt)$ channels, at $y_{uu}=0.2$.}
\label{fig:p-val-chan}
\end{figure}

For the additional decay channels, the sensitivity is weaker than for $\chi \to Wb$, with local $p$-values reaching only about $2\sigma$ for $\chi\to Zt$ and $3\sigma$ for the $\chi\to h^0t$, as shown in \Cref{fig:p-val-chan}.

For the discovery channel, we evaluated the significance using both toy-based and simple Poisson statistics. In most of the $M_S$ range, the toy studies yield $p$-values consistent with zero, in line with the extremely small probabilities obtained from Poisson mass function. Such vanishing $p$-values for large Yukawa couplings should be regarded as unphysical artifacts of the idealized setup. 

Overall, the limits in the fully hadronic $u\chi$ channel are only weakly dependent on $m_\chi$ and $y_{uu}$, but improve significantly for larger $y_{u\chi}$. The $\chi\to Wb$ mode dominates the sensitivity, while $\chi\to Zt$ and $\chi\to h^0t$ are less impactful.

\section{Conclusions}\label{sec:conclusions}

With the High Luminosity LHC upgrade, the more than one order-of-magnitude increase in integrated luminosity will substantially enhance the reach for high-mass resonances. In this context, diquark scalars and vectorlike quarks have attracted renewed interest, especially following the CMS and ATLAS reports of high-invariant-mass multijet events \cite{ATLAS:2024gyc, CMS:2022fck}.

In this work we investigate a new decay channel $S_{uu}\to u\chi$, whose cross section is smaller than the $S_{uu}\to \chi\chi$ decay mode previously analyzed in \cite{Duminica_2025, Costache_2025}. We use the same simulation and analysis frameworks to generate and separate signal from background processes. Fully hadronic final-state cross sections and signal-to-background selection efficiencies are compared for two vectorlike-quarks masses, three benchmark choices of Yukawa couplings, and different combinations of the three $\chi$ decay modes.

Lowering the vectorlike-quark mass from 2 to 1.5 TeV only has a small effect: the sampled signal yield decreases slightly with $m_\chi$, in agreement with the theoretical expectations in  \cite{Dobrescu:2018psr}. Among the three Yukawa benchmarks, the scenario with enhanced $y_{u\chi}$ provides the best sensitivity in the $u\chi$ channel. For all considered $\chi$ decay modes, the Random Forrest classifier achives good discrimination performance, reinforcing the motivation for a future combined analysis including all multijet final-state channels.  

In the fully hadronic $S_{uu}\to u \chi \to Wb$ channel, the 95\% C.L. exclusion limits are around $M^{lim}_S\simeq 8.2$ TeV for $m_\chi=1.5,~2$ TeV and $y_{uu}B(\chi\to Wb) = 0.1, \ 0.2$, with a discriminator threshold $D=0.9$. The mass reach is reduced for the other $\chi$ decay modes, yielding $M^{lim}_S\approx 7.4$ TeV for $\chi\to Zt$, and $M_{S}^{lim}\approx 7.9$ TeV when $\chi\to h^0t$. In the benchmark with enhanced $y_{u\chi}$, the 95\% C.L. limit improves to $M_{S}^{lim}\to 9.2$ TeV. Compared with the $S_{uu}\to\chi\chi$ channel, the Random-Forrest-based and statistical analyses are less sensitive in the $u\chi$ case, mainly due to the smaller production cross section and correspondingly lower signal yields.

Although we adopt a different theoretical benchmark, the exclusion limits obtained in this work are broadly consistent with the CMS search for resonantly produced dijet pairs via broad mediators \cite{CMS:2025hpa}, which reports four-jet invariant-mass limits of $8$ TeV for narrow and $9$ TeV for broader resonances. Since this is purely a phenomenological study, our statistical results should be understood as estimates of the sensitivity of the model to the diquark and vectorlike quark masses, rather than as definitive experimental exclusions.

Local $p$-values range between $\sim 1\sigma$ and $\sim 4\sigma$, depending on the decay mode, Yukawa couplings, $M_S$ and $m_\chi$, indicating that the phenomenology of this channel play an important role in providing evidence for ultraheavy diquarks at the HL-LHC.

A natural next step is to embed this channel in a global program of diquark searches, combining the fully hadronic $u\chi$ topology with the previously studied $\chi\chi$ and $uu$ final states in a unified likelihood. Such a combined analysis, incorporating realistic detector effects and systematic uncertainties, would provide more robust constraints on the $S_{uu}$ mass and its Yukawa couplings. On the theory side, extending the study to include NLO QCD corrections, PDF uncertainties at large x, and alternative flavor structures for the $S_{uu}$ couplings would help quantify the model dependence of our results. Together, these developments would sharpen the role of ultraheavy diquark scalars and vectorlike quark searches as benchmark targets for the HL-LHC multijet physics program.

\section*{Acknowledgments}
The work of M.-S.F., C.A., D.-C.C., I.-M.D. and I.D. was supported by IFIN-HH under Contract No.~PN-23210104 with the Romanian Ministry of Education and Research. The work of G.M. was supported by the project "Romanian Hub for Artificial Intelligence - HRIA", Smart Growth, Digitization and Financial Instruments Program, 2021-2027, MySMIS no.~334906.

\providecommand{\noopsort}[1]{}\providecommand{\singleletter}[1]{#1}%

\end{document}